\documentstyle[12pt,psfig]{article}
\begin{document}

\title{Nucleus-nucleus collisions at high baryon densities}
\author{H.~Weber, E.~L.~Bratkovskaya\thanks{Supported by DFG}\hspace{5pt}
and H.~St\"ocker\\[2mm]
{\normalsize Institut f\"{u}r Theoretische Physik,
Universit\"{a}t Frankfurt}\\
{\normalsize 60054 Frankfurt, Germany}}
\date{ }
\maketitle

\begin{abstract}
We study central collision of $Pb + Pb$ at 20, 40, 80 and 160
A$\cdot$GeV within the UrQMD transport approach and compare rapidity
distributions of $\pi^-, K^+, K^-$ and $\Lambda$ with the recent
measurements from the NA49 Collaboration at 40, 80 and 160 A$\cdot$GeV.
It is found that the UrQMD model reasonably describes the data,
however, systematically overpredicts the $\pi^-$ yield by $\sim 20$\%,
whereas the $K^+$ yield is underestimated by $\sim 15$\%.
The $K^-$ yields are in a good agreement with the experimental data,
the $\Lambda$ yields are also in a reasonable correspondence with the
data for all energies. We find that hadronic flavour exchange reactions
largely distort the information about the initial strangeness
production mechanism at all energies considered.
\end{abstract}

\vspace{0.3cm}\noindent


\vspace*{10mm}
%---------------------------------------------------------------------
\newpage

The dynamics of nucleus-nucleus collisions at high baryon density --
contrary to the high energy density at RHIC and LHC -- is of present
interest with respect to the parton/hadron phase transition at high
quark chemical potential. Furthermore, as has been proposed early by
Rafelski and M\"uller \cite{Rafelski} the strangeness degree of freedom
might play an important role in distinguishing hadronic and partonic
dynamics. This also relates to the entropy per baryon (or number of
constituent quarks) which provides information on the effective number
of degrees of freedom involved. Additionally, one expects that at high
net quark density the chiral symmetry of QCD -- which is broken in the
vacuum as reflected in the non vanishing quark condensate
$\langle\bar{q}q\rangle$ -- becomes restored for considerable
space-time intervals (cf.~Figs.~3,4 in \cite{HSD_exf}). Thus
the (at least partial) restoration of chiral symmetry should lead
to dramatic changes of the hadron spectral functions, which either
might show poles at zero masses \cite{BrownRho} or a complete mixing
for chiral partners such as the $\rho$- and $a_1$-mesons
\cite{Zahed,KoKoch}. The related questions are addressed in more detail
in the proposal for the (discussed) future heavy-ion facility at GSI
Darmstadt \cite{GSIprop}.

Whereas experimentally the dynamics of heavy nucleus-nucleus collisions
have been studied up to 11.6 A$\cdot$GeV at the BNL AGS and an
extensive program has been carried out at the 'top' CERN SPS energy of
160 A$\cdot$GeV, the intermediate range from $\sim 11$ to 160
A$\cdot$GeV essentially has been 'terra incognita'.  Only recently,
experiments have been carried out at the CERN SPS for 40 and 80
A$\cdot$GeV \cite{NA49_new,NA49_Lam} and further experimental studies
are foreseen at 20 A$\cdot$GeV and 30 A$\cdot$GeV \cite{SPS20}.  The
elementary question thus arises to what extend we might find signatures
for an intermediate QGP state or do we just see strongly interacting
hadronic matter \cite{Ehehalt,Soff:1999et,Bass:1997xw} ?

The experimentally measured $K^+/\pi^+$ ratio, which has been suggested
to be a deconfinement indicator (see e.g.\ \cite{NA49_new,Marek} and
references therein), shows a non-monotonic  behaviour with a possible
maximum between 11.6 and 40 A$\cdot$GeV. Such behaviour  can not be
fully reproduced by different microscopic transport approaches as RQMD
\cite{RQMD}, HSD \cite{HSD_exf} and UrQMD \cite{Weber_ratio} (for more
details see e.g.\ \cite{Bass} and references therein). The statistical
model \cite{StatMod} is in better agreement with the data, however, it
can not shed light on  chiral symmetry restoration or the existence of
a QGP since this model is based on hadronic degrees of freedom, which
are even non-interacting (contrary to the dynamical picture of
transport models).  The failure of transport approaches -- based on
hadronic and quark-string degrees of freedom \cite{Weber:1998zb} -- to
reproduce the $K^+/\pi^+$ ratio has been interpreted in \cite{HSD_exf}
as a possible indication for the formation of unbound quark matter at
high baryon density reached in the initial phase of central $Au + Au$
(or $Pb + Pb$) collisions.

Whereas the $K^+/\pi^+$ ratio is basically attributed to the
midrapidity yields, it is important to look also at the full rapidity
range and independently on pion and strange particle yields for a
better understanding of the collisional dynamics. The recent high
accuracy data from the NA49 Collaboration \cite{NA49_new,NA49_Lam}
allow to make a more conclusive comparison on this issue.  In this work
we study central nucleus-nucleus collisions within a microscopic
transport approach -- the ultra-relativistic quantum molecular dynamics
(UrQMD) model (version 1.3) -- and compare with the data from the NA49 Collaboration
for central $Pb + Pb$ collisions at 40, 80 and 160 A$\cdot$GeV. Also we
make predictions for pion and strangeness production at 20 A$\cdot$GeV.

The UrQMD transport approach is described in Refs.~\cite{UrQMD1,UrQMD2} 
and includes all baryonic resonances up to an invariant mass of 2 GeV 
as well as mesonic resonances up to 1.9 GeV as tabulated in the PDG 
\cite{PDG}. For hadronic continuum excitations we employ a string model 
with meson formation times in the order of 1-2~fm/c depending on the 
momentum and energy of the created hadrons.  The transport approach is 
matched to reproduce the total nucleon-nucleon, meson-nucleon and 
meson-meson cross section data in a wide kinematical regime 
\cite{UrQMD1,UrQMD2}. We note, that  uncertainties remain with respect 
to the differential spectra in rapidity $y$ and transverse momentum 
$p_T$, that are not sufficiently controlled by experimental data 
especially when short-lived resonance states are involved in the 
reaction. At the high energies considered here the particles are 
essentially produced in primary high-energy collisions by string 
excitation and decay, however, the secondary interactions among 
produced particles (e.g.\ pions, nucleons and excited baryonic and 
mesonic resonances) also contribute to the particle dynamics -- in  
production as well as in absorption. In transport calculations all 
global symmetries like baryon number, charge and strangeness are 
strictly conserved as well as energy and momentum in each individual 
reaction.

\begin{figure}[ht]
\centerline{\psfig{figure=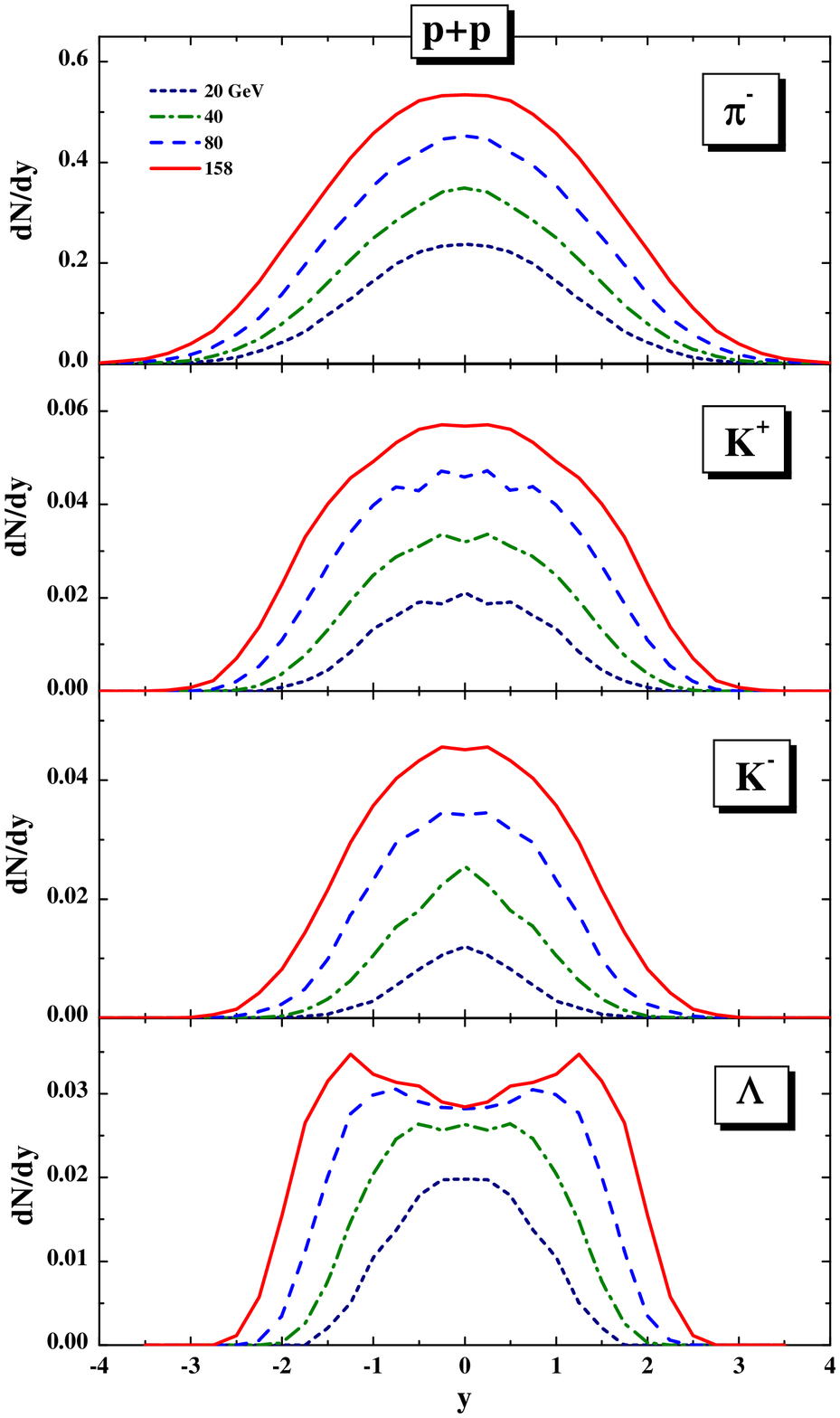,width=12cm}}
\caption{The rapidity distribution of $\pi^-, K^+, K^-$ and
$\Lambda (+\Sigma^0)$ particles in $pp$  collisions at 20, 40, 80
and 160 GeV calculated within the UrQMD model.}
\label{urqmdpp}
\end{figure}

Before coming to the results for central nucleus-nucleus collisions it
is instructive to  look at the UrQMD results for $\pi^-, K^\pm$ and
$\Lambda (+\Sigma^0)$ rapidity spectra from $pp$ collisions at the same
bombarding energies per nucleon. The calculated rapidity spectra
(normalized to the total $pp$ cross sections) are shown in Fig.
\ref{urqmdpp} for $\pi^-, K^\pm$ and $\Lambda$'s at 20, 40, 80 and 160
GeV, respectively. One observes a smooth increase with energy of the
$\pi^-$ and $K^+$ spectra, both in magnitude and width. This increase
with energy is more pronounced for the antikaons, which show a
significantly smaller width in rapidity than the $K^+$ mesons. On the
other hand, the $\Lambda (+\Sigma^0)$ midrapidity spectra are almost
constant from 40 - 160 GeV while the width in rapidity increases
substantially with energy.  These general tendencies have to be kept in
mind when interpreting the calculated results from central
nucleus-nucleus collisions (see below).  Furthermore, related
differential experimental spectra would be highly welcome to shed some
light on the dominant 'elementary' production process of pions and
strange hadrons at these energies.

We continue with the related results for pions, kaons, antikaons and
hyperons from $A+A$ collisions and start at the highest bombarding
energy of 160 A$\cdot$GeV. A comparison of our calculations for the
most central (5\%) $Pb + Pb$ collisions at 160 A$\cdot$GeV with the
data from Ref.~\cite{NA49_new,NA49_Lam} is shown in Fig.~2 for $\pi^-,
K^+, K^-$ and $\Lambda (+\Sigma^0)$.  We note that the centrality of
the reactions has been determined by a comparison of our calculations
to the energy distribution in the experimental Veto-calorimeter from
NA49.  It is seen that the spectral shape is rather well reproduced,
however, the $\pi^-$ yield is overestimated by $\sim 17$\%, whereas the
$K^+$ and $K^-$ spectra are underpredicted by $\sim 15$\% and $\sim
6$\%, respectively. The $\Lambda (+\Sigma^0)$ rapidity distribution is
on the upper level of the experimental error bars.  Though it has been
claimed in Ref.~\cite{Heinz}, that a QGP might have been seen, the
comparison of the hadron resonance/string approach with the data on
$\pi^-, K^\pm, \Lambda (+\Sigma^0)$ rapidity distributions does not
show clear signs of new (i.e.~partonic) degrees of freedom at this
energy.  This finding essentially agrees with independent studies in
the HSD transport approach \cite{HSD_K,CBRep98}.

We now step down in energy to the intermediate regime that has not been
investigated experimentally so far. The data of the NA49 Collaboration
for $Pb + Pb$ at 80 A$\cdot$GeV \cite{NA49_new} for $\pi^-, K^+, K-$
and $\Lambda (+\Sigma^0)$ are shown in Fig.~2 (3rd column) and compared
to our calculations for the central (7\%) events. Here we again observe
an overestimation of the $\pi^-$ yield by $\sim 20$\%, a very good
description of the $K^-$ rapidity spectra and a reasonable agreement
with the data for the $\Lambda (+\Sigma^0)$ rapidity distribution.  The
$K^+$ yield falls off in the calculation by $\sim 17$\% such that the
experimental $K^+/\pi^+$ ratio is underestimated by $\sim 30$\% from
the UrQMD calculations.  The situation is similar at 40 A$\cdot$GeV for
the central (7\%) collisions of $Pb + Pb$ (cf.~Fig.~2 - 2nd column)
where $K^-$ and $\Lambda (+\Sigma^0)$ rapidity distributions are well
described, the $K^+$ yield is underestimated by $\sim 15$\% while the
$\pi^-$ spectrum from the calculations is too high by about  $\sim
25$\%.

We note in passing that a simple strangeness counting rule, i.e.
$N_\Lambda +N_{K^-}\simeq N_{K^+}$, does not hold in our case since
the $\Lambda$ yields include the decay from $\Sigma^0$ and in the total
strangeness balance also $K^0, \bar\Lambda, \bar\Sigma$ etc. (with $K^+$)
and $\Sigma^\pm$ as well as $\Xi$'s and $\Omega$'s have to be
considered, too.

Our predictions for the 7\% most central $Pb + Pb$ collisions at 20
A$\cdot$GeV, that will be measured at the SPS \cite{SPS20} and possibly
in more detail at the future GSI facility \cite{GSIprop}, are shown in
Fig.~2 (1st column) for  $\pi^-, K^+, K^-$ and $\Lambda (+\Sigma^0)$.
Following the trend from the higher energies in Fig. 2 we expect also
to overpredict the $\pi^-$ yield and to underestimate the $K^+$ cross
section.

The question remains to what extend the deviation of our transport
calculations from the data in Fig. 2 might indicate new physics or the
traces from partonic degrees of freedom. To this aim in
Fig.~\ref{chandec} we present the channel decomposition (fraction in
\%) for the final $K^+$ (upper part) and $K^-$ yields (lower part)
calculated for central ($b=0$~fm) $Pb + Pb$ collisions at 20, 40, 80
and 160 A$\cdot$GeV. In order to explain the results from
Fig.~\ref{chandec} we note that initially $s, \bar s$ quarks are
produced in high energy nucleon-nucleon collisions and later on in
meson-baryon interactions via string excitations and decays.  However,
afterwards the strange particles (produced initially) participate in
chemical reactions with flavor exchanges. Thus only a few percent of
the 'primary' kaons/antikaons remain unaffected by secondary inelastic
interactions (cf.~the lines denoted as 'BB string' in
Fig.~\ref{chandec}).  Most of the final $K^+$ and $K^-$ mesons finally
stem from $K^{*\pm}(892)$ decays (lines '$K^*$ decay') which are either
produced directly in string decays or by pion-kaon resonant scattering.
About 2\% of the final $K^+$ and $\sim$5\% of $K^-$ appear from the
$\phi(1020)$ meson decays (lines '$\phi$ decay'). The lines '$m^*$
decays' denote the fraction of final kaons and antikaons coming from
higher mesonic resonance decays (i.e.~$K^*(1410), K^*(1680),
K_0^+(1430), a_0(980), f_0(980)$, etc.). At the SPS energies considered
here only a small fraction of the final kaons/antikaons can be
attributed to baryonic resonance decays ('$B^*$ decays'). This fraction
slightly increases when lowering the bombarding energy. About 15-20\%
of $K^+$ and $~20$\% of $K^-$ stem from meson-baryon string decays
('$mB$ string') excited in energetic secondary meson-baryon
interactions that do no longer participate in further inelastic
reactions. Note, that in the channel denoted as '$mB$ string' the
kaon/antikaon-baryon collisions are also counted.

The picture, which emerges from the interpretation of Fig.
\ref{chandec}, thus is as follows: only a small fraction of
kaons/antikaons from energetic initial collisions survives the hadronic
rescattering phase during the expansion of the fireball. Most of the
final strangeness yield emerges after rescattering -- shifting $s$
quarks from mesons to baryons and vice versa -- thus providing a very
distorted picture on the initial strangeness production mechanism and
the elementary degrees of freedom involved. Thus the $K^\pm$ and
$\Lambda (+\Sigma^0)$ spectra do not allow for stringent conclusions on
the initial phase of high energy density. On the other hand, these
frequent flavor exchange reactions might be the reason why statistical
models - employing chemical equilibration - seem to work reasonably
well.

In summary, our detailed transport study with the UrQMD approach for
central collisions of $Pb + Pb$  at 20, 40, 80 and 160 A$\cdot$GeV has
shown that the UrQMD model -- involving string as well as hadronic
degrees of freedom -- reasonably describes the data from the NA49
Collaboration, however, systematically overpredicts the $\pi^-$ yield
by $\sim 25$\%. On the other hand, the $K^+$ yield is underestimated by
$\sim 15-20$\% from 40 and 80 A$\cdot$GeV while the $K^-$ yields are in
a good agreement with the data for all energies. The $\Lambda
(+\Sigma^0)$ multiplicities are found to roughly reproduce the data for
all energies.  The explicit channel decomposition of the final $K^\pm$
suggests that the kaon/antikaon rapidity spectra do not allow to
determine the effective degrees of freedom -- either partonic or
string/hadron like -- in the initial phase of the reaction due to the
strong hadronic interactions in the expansion phase of the system.

The systematic overprediction of pions and underprediction of $K^+$
mesons might suggest that the hadron/string approach systematically
fails in the energy regime from 20-160 A$\cdot$GeV especially when
looking at the $K^+/\pi^+$ ratio (cf. Refs.  \cite{HSD_K,CBRep98}).
Such a 'failure' might indicate the presence of partonic degrees of
freedom in the initial phase of the collision and/or reflect a partial
restoration of chiral symmetry \cite{HSD_exf}. However, some cautious
remarks appear necessary: presently it is not clear if all the
differential hadronic reactions employed in the transport calculation
are sufficiently controlled by experimental data. This is even obvious
for reactions involving short-lived resonance states. Thus it might
well be that the $\leq $20\% differences found in comparison to the
NA49 hadron spectra could be attributed to uncertainties in hadronic
cross sections or string fragmentation functions. Some further
theoretical work and related data on the 'primary' $NN$ and $\pi N$
reactions will be necessary to clarify this presently open issue.

\vspace*{5mm}
The authors are grateful to M.~Ga\'zdzicki and T. Kollegger for providing
us with the data from the NA49 Collaboration in digital form.

%------------------------------------------------------------------------

%--------------------------------------------------------------
\newpage

\begin{figure}[ht]
\centerline{\psfig{figure=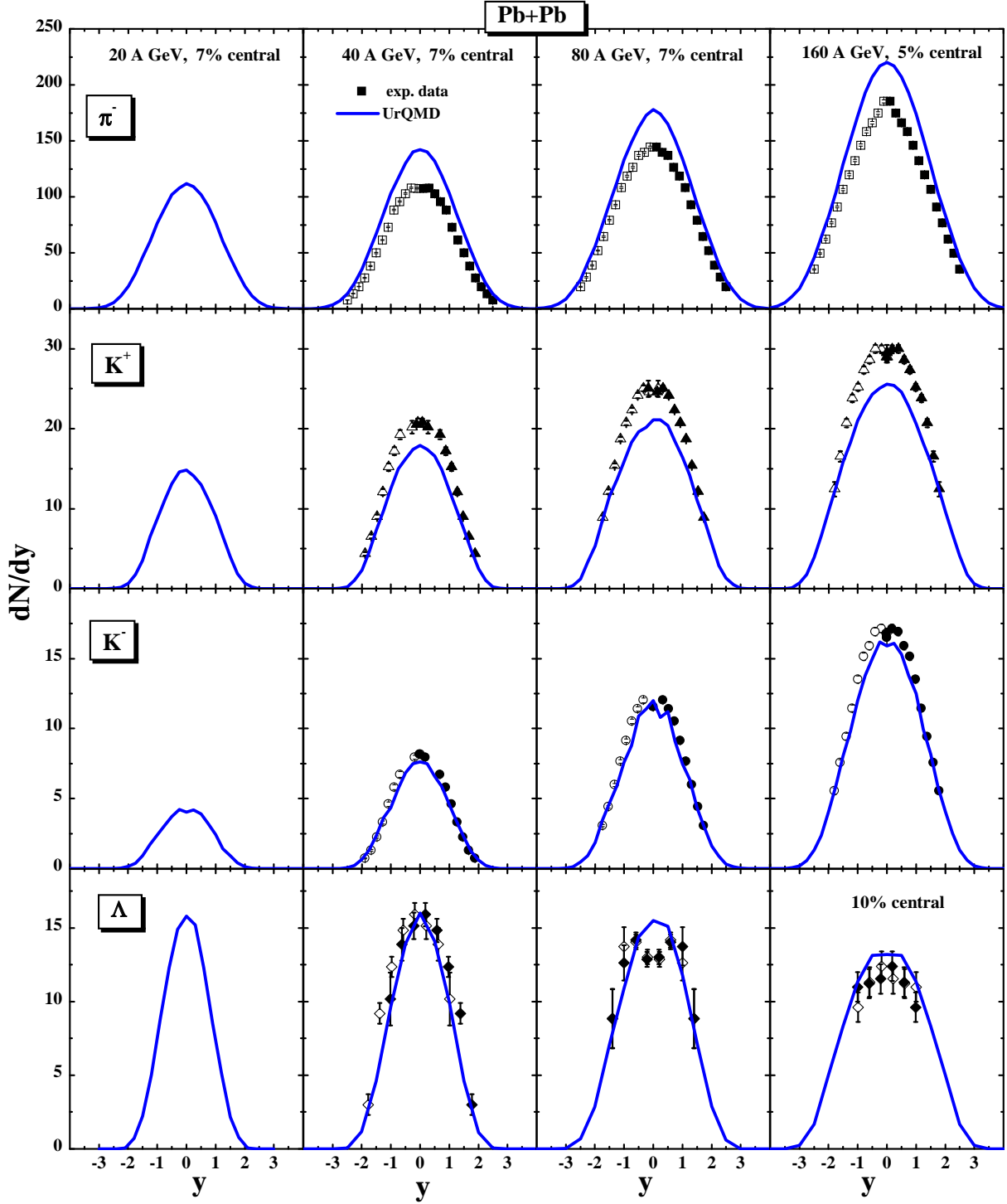,width=16cm}}
%\vspace*{-5mm}
\caption{The rapidity distribution of $\pi^-, K^+, K^-$ and $\Lambda
(+\Sigma^0)$ in 7\% or 5\% central $Pb + Pb$ collisions at 20, 40, 80
and 160 A$\cdot$GeV calculated within the UrQMD model (solid lines) in
comparison to the experimental data from the NA49 Collaboration at 40,
80 and 160 A$\cdot$GeV:  the squares represent $\pi^-$, triangles --
$K^+$, circles -- $K^-$ \cite{NA49_new} and the diamonds indicate
$\Lambda (+\Sigma^0)$ experimental data \cite{NA49_Lam}.  The full
symbols correspond to the measured data, whereas the open symbols are
the data reflected at midrapidity. Note, that the $\Lambda (+\Sigma^0)$
experimental data (as well as UrQMD results) at 160 A$\cdot$GeV
correspond to 10\% central $Pb + Pb$ collisions.}
\label{urqmd160}
\end{figure}

\begin{figure}[ht]
\centerline{\psfig{figure=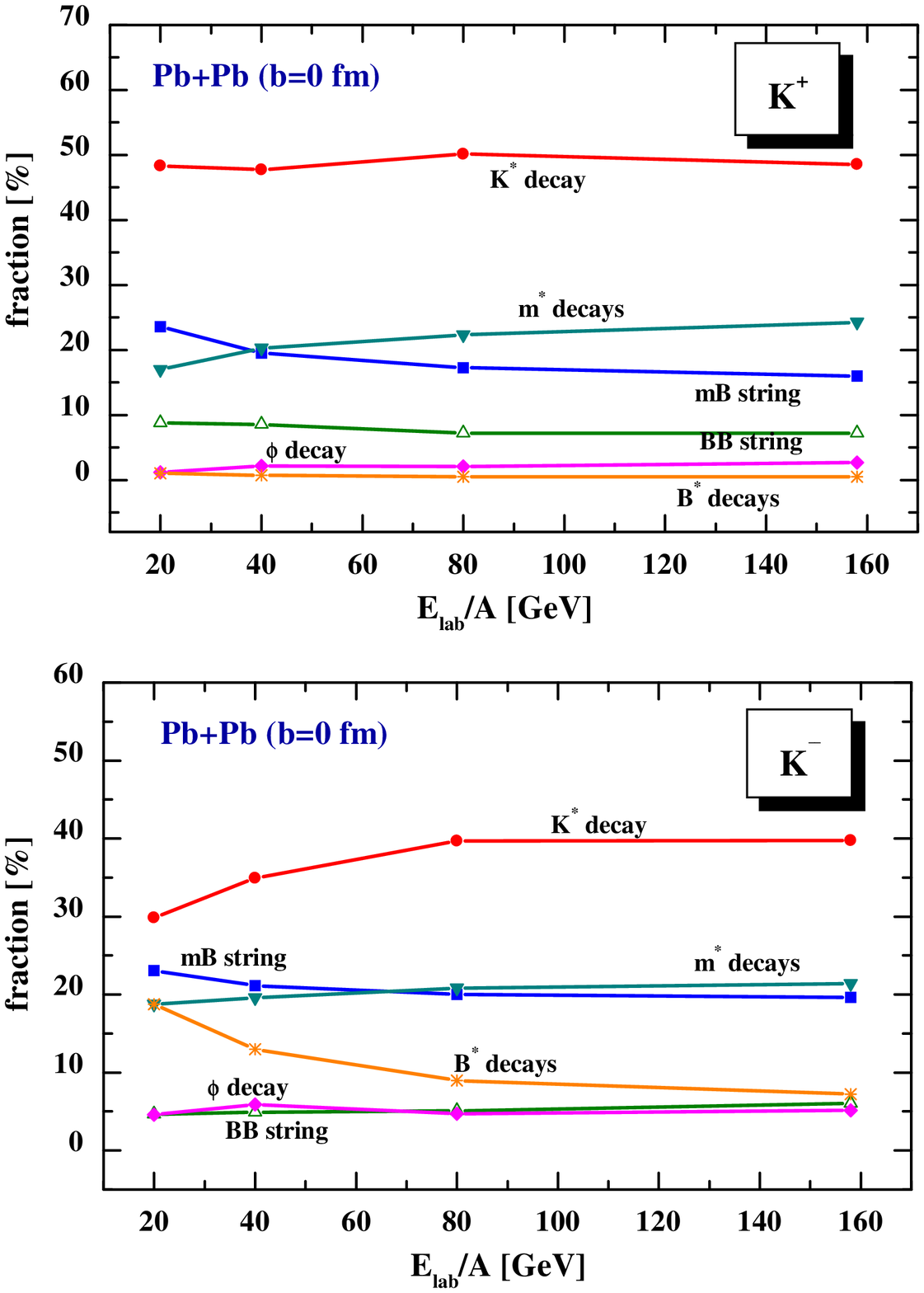,width=15cm}}
\caption{Channel decomposition for the final $K^+$ (upper part) and
$K^-$ yields (lower part) calculated within UrQMD for central
($b=0$~fm) $Pb + Pb$ collisions at 20, 40, 80 and 160 A$\cdot$GeV.}
\label{chandec}
\end{figure}

\end{document}